\documentclass[final]{svjour2}
\usepackage{graphicx}
\usepackage{rotating}
\usepackage{amssymb}
\usepackage{mathptmx}
\usepackage[numbers]{natbib}
\makeatletter
\journalname{Journal of Low Temperature Physics}
\bibpunct{}{}{,}{s}{}{,}

\begin{document}

\newcommand{\hdblarrow}{H\makebox[0.9ex][l]{$\downdownarrows$}-}
\title{Transport in Fermi Liquids Confined by Rough Walls}

\author{Priya Sharma}

\institute{Theoretical Sciences Unit, Jawaharlal Nehru Centre for Advanced Scientific Research, Jakkur, Bangalore 560064, India.}
%Tel.:\\ Fax:\\
%\email{priyasharma@jncasr.ac.in}
%\\2: XXXXXXXXX\\ 3: YYYYYYYYYYY}

\date{01.07.2013}

\maketitle

\begin{abstract}
I present theoretical calculations of the thermal conductivity of Fermi liquid $^3$He confined to a slab of thickness of order $\sim 100$nm. The effect of the roughness of the confining surfaces is included directly in terms of the surface roughness power spectrum which may be determined experimentally. Transport at low temperatures is limited by scattering off rough surfaces and evolves into the known high-temperature limit in bulk through an anomalous regime in which both inelastic quasiparticle scattering and elastic scattering off the rough surface coexist. I show preliminary calculations for the coefficients of thermal conductivity. These studies are applicable in the context of electrical transport in metal nanowires as well as experiments that probe the superfluid phase diagram of liquid $^3$He in a slab geometry.

%PACS numbers: 74.70.Tx,74.25.Ha,75.20.Hr
\end{abstract}

\section{Introduction}

Quantum size effects arise in Fermi liquids when quasiparticles are confined to restricted geometries. The nature of the confining walls sets the boundary conditions for the interaction of quasiparticles by scattering off the surface of the wall. When the surface of the boundary is rough, quasiparticles scatter off the rough surface in addition to scattering off one another. This modifies transport properties in the Fermi liquid as there is an additional relaxation channel available to quasiparticles. In this paper, I consider thermal transport in a Fermi liquid confined to a slab with rough walls. I calculate the effect of the coexistence of various scattering channels on the transport of heat in this system. The thermal conductivity is an important response coefficient in the characterization of films of $^3$He used in experiments aimed at studying the superfluid phase diagram of a $^3$He film{\cite{scienceSlab}}. It is also technologically important in understanding transport in the Fermi liquid of electrons in thin metallic nanowires.

\section{Theory}
In thin films, the spatial confinement leads to quantization of quasiparticle momenta perpendicular to the walls. The quasiparticle energies $\epsilon(\vec{p})$ split into a set of minibands $\epsilon_j(p_{\parallel})$, where $p_{\parallel}$ is the component of quasiparticle momentum parallel to the walls. When the scattering surface is rough with roughness being slight viz., the characteristic length scale associated with the roughness $\ell\ll L$, where $L$ is the thickness of the slab; the quantization due to confinement can be treated as local and is hence spatially varying. For slabs with thickness much larger than the inverse Fermi wavelength, $k_F L \gg 1$ and slight surface roughness, various scattering channels can coexist - bulk inelastic binary quasiparticle scattering, elastic scattering from impurities and scattering off the rough surface. While the boundary condition is usually imposed as a constraint on quasiparticles scattering in various channels, an alternate approach proposed by Meyerovich and coworkers{\cite{Mey98,Mey01,Mey02}} incorporates the boundary condition in a coherent fashion. This treatment is robust to the (coherent) coexistence of more than one scattering mechanism and hence incorporates potential interference effects as well as correlations between scattering centres in a particular regime defined by external parameters. A mapping transformation is applied{\cite{Tesanovic,Trivedi}} to map the surface scattering to a virtual disorder potential that operates in the bulk, mapping  the geometry to one with flat walls. Averaging over the impurities and surface roughness, the self-energy terms are built including the three types of interactions viz., inelastic quasiparticle scattering, elastic scattering off impurities and elastic scattering off the virtual disorder potential derived from the surface roughness. The imaginary part of the self-energy gives a wall-induced transition probability, $W_{jj'}(\bf{q},\bf{q'})$ for ballistic quasiparticles scattering off the disorder potential mediated by binary quasiparticle scattering processes. The relaxation rate for a slab with one rough wall, in the absence of bulk disorder is given by{\cite{Mey01}},
\begin{eqnarray}
\label{MeyerovichScatteringRate}
\nonumber
\frac{1}{\tau_j^{eff}(\vec{p})} &=& \frac{1}{\tau_j^{(b)}(\vec{p})} + \sum_{j'=1}^{S}\int\nolimits \frac{W_{jj'}({\bf{q}},{\bf{q'}})/\tau_{j'}^{(b)}(\bf{q'})}{(\epsilon_{j'}({\bf{q'}})-\epsilon_F)^2/\hbar^2 + (1/2\tau_{j'}^{(b)}({\bf{q'}}))^2} \frac{d{\bf{q'}}}{(2\pi\hbar)^2}\,,\\
W_{jj'}({\bf{q}},{\bf{q'}}) &=& \frac{\pi^4\hbar^2}{M^2L^6}\zeta({\bf{q}}-{\bf{q')}}j^2 j'^2\,\,\,
\end{eqnarray}
where $1/\tau_j^{(b)}$ is the bulk relaxation rate for inelastic quasiparticle scattering in band indexed by $j$, $M$ is the quasiparticle effective mass and $\zeta$ is the autocorrelation function of the surface roughness of the slab wall. This effective rate has been used to calculate the relaxation of a film of $^3$He in a torsional oscillator{\cite{PRL,QFS2012JLTP}}with the surface roughness power spectrum used as an independent input parameter. The calculation showed good agreement with experimental results, including a direct input of AFM data determined for the surface used. In this paper, I calculate thermal transport for a slab with one rough surface whose roughness is set as an independent input.

The thermal conductivity can be calculated from the Landau-Boltzmann equation for the quasiparticle distribution function, $n_{\vec{p}\sigma}$. The driving term of this transport equation is the collision integral which gives the rate of change of the distribution function as a result of quasiparticle collisions. The effective scattering rate given by equation({\ref{MeyerovichScatteringRate}}) is set as the driving term for the transport equation in this case.

It is sufficient to consider linear deviations from thermal equilibrium and the linearized Boltzmann equation in the steady state for this case is
\begin{equation}
-\frac{\partial n_{\vec{p}\sigma}^0}{\partial \epsilon_{\vec{p}\sigma}}\,\vec{v_{\vec{p}\sigma}}\cdot\nabla(\delta\epsilon_{\vec{p}\sigma}) = I[n_{\vec{p}\sigma}]\,\,\,,
\end{equation}
where $n_{\vec{p}\sigma}^0$, $\epsilon_{\vec{p}\sigma}$ and $\vec{v_{\vec{p}\sigma}}$ are the equilibrium distribution function, energy and velocity of quasiparticles of momentum $\vec{p}$ and spin $\sigma$ respectively. $I[n_{\vec{p}\sigma}]$ is the collision integral which is a functional of the distribution function, $n_{\vec{p}\sigma}$. In general, the collision integral includes all forms of scattering viz., pure inelastic scattering processes of quasiparticles off one another as in the bulk, elastic scattering off the boundary which has been included here as a virtual disorder potential in the bulk, as well as mixed scattering processes which are the latter mediated by inelastic scattering, to leading order in the surface roughness $\xi/L$ which is treated as a perturbative parameter. As a first step, we consider the case with a collision integral that includes elastic scattering off the virtual disorder potential mediated by binary quasiparticle scattering processes. The rate of change of the distribution function is given by linearizing the scattering rate in equation(\ref{MeyerovichScatteringRate}). For slab thicknesses with $k_F L \gg 1$, transport is quasiclassical and quasiparticles may be regarded as travelling along well-defined trajectories with momentum $p\sim p_F$. In this quasiclassical limit, the summation over the band index $j$ is replaced by integration over the continuous variable, $\pi j\hbar/L\rightarrow p_z$, where $z$ is the direction of confinement i.e, the direction perpendicular to the walls. The transport equation can then be solved for the deviation of the quasiparticle distribution function from equilibrium, $\delta n_{\vec{p}\sigma} = n_{\vec{p}\sigma} - n_{\vec{p}\sigma}^0$. 

\section{Preliminary Results}

The heat current, $\vec{j}$ is given by
\begin{equation}
\vec{j} = 2\int\nolimits \frac{d^3p}{(2\pi\hbar)^3}\,(\epsilon_{\vec{p}}-\mu) \vec{v_{\vec{p}}}\delta n_{\vec{p}} = \bar{\kappa} \nabla T\,\,\,,
\end{equation}
where the factor $2$ is for the spin sum, $\mu$ is the chemical potential and $\bar{\kappa}$ is the thermal conductivity tensor in response to the thermal gradient $\nabla T$. I calculate
\begin{equation}
\label{kappa}
\kappa_{xx} = {\kappa_0}\int_{-\infty}^{\infty} dx\, x^2\, sech^2(\frac{x}{2})(x^2 {\frac{k_B^2 T^2}{\hbar^2}} + (\frac{1}{2\tau_b})^2)\int_0^{2\pi}\int_0^{\pi} \frac{d(cos\theta) d\phi}{4\pi} tan^2\theta cos^2\phi\,\,S(\theta,\phi),
\end{equation} 
where
\begin{equation}
\kappa_0 = \frac{k_B\hbar^2 L \tau_b}{4 p_F^2}\,\,\,,
\end{equation}
$\tau_b = \tau_0/T^2$ is the bulk scattering time and $\tau_0$ is well-known for bulk $^3$He. $\theta$ and $\phi$ are the polar and azimuthal angles of the quasiparticle momentum, $\vec{p}$ with respect to the direction of confinement, $z$. The function $S(\theta,\phi)$ involves the surface roughness power spectrum $\zeta$,
\begin{equation}
S(\theta,\phi) = \int_0^{2\pi}\int_0^{\pi} \frac{d(cos \theta') d\phi'}{4\pi} \zeta(\vec{q}-\vec{q'})\,cos^2\theta'\,(1 - \frac{sin\theta'\,cos\phi'}{sin\theta\,cos\phi})\,\,\,.
\end{equation}
Here $\vec{q}\equiv p_{\parallel}$. Similar expressions can be derived for $\kappa_{xy}$, $\kappa_{xz}$, etc. The various components of the $\kappa$ tensor are given by expressions that differ only in the angular integrals and show the same temperature dependence. Fig.1 shows $\kappa_{xx}$ as a function of temperature and film thickness for a surface with Gaussian surface roughness given by $\zeta(\vec{q}) = 2\pi l^2 R^2 e^{-q^2 R^2/2}$.
\begin{figure}
\begin{center}
\includegraphics[%
  width=0.65\linewidth,
  keepaspectratio]{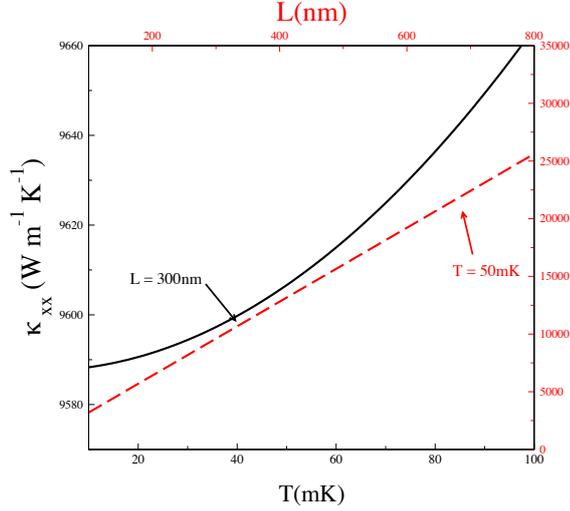}
\end{center}
\caption{(Color online) $\kappa_{xx}$ as a function of temperature (solid) for a film thickness of $L=300$nm with $\kappa_{xx}$-axis on the left;  and as a function of film thickness (dashed) at $T=50$mK with $\kappa_{xx}$-axis on the right. Surface roughness parameters $l=10$nm and $R=50$nm.}
\end{figure}

\section{Discussion}

The thermal conductivity, $\kappa_{xx}$ in equation(\ref{kappa}) consists of two terms. The constant temperature-independent term is larger at lower temperatures and gives the residual thermal conductivity originating from wall-scattering (which is independent of temperature and given by the surface roughness). The second term in the thermal conductivity has a quadratic temperature dependence and arises from the interplay of both inelastic and elastic scattering mechanisms that enable effective heat transport by quasiparticles.

The effect of surface roughness in a Fermi liquid confined to a geometry where quantum size effects can be important is nontrivial and gives rise to anomalous transport properties over a range of temperatures. This anomalous behaviour should be observable in nanoscale cavities filled with liquid $^3$He.

\begin{acknowledgements}
I would like to thank the Jawaharlal Nehru Centre for Advanced Scientific Research, Bangalore for hosting me during the course of this work. I am grateful to M. Eschrig for several helpful discusssions, and J. Saunders for his support.
\end{acknowledgements}

%\pagebreak

\end{document}